\documentclass[%
aip,
 amsmath,amssymb,
 reprint,%
]{revtex4-1}

\usepackage{graphicx}
\usepackage{dcolumn}
\usepackage{bm}

\usepackage[utf8]{inputenc}
\usepackage[T1]{fontenc}
\usepackage{mathptmx}

\begin{document}

\preprint{AIP/123-QED}

\title{Novel method for determination of contact angle of highly volatile liquids}

\author{K Nilavarasi}
\email{nilavarasikv@gmail.com}
\author{V Madhurima}%
\email{madhurima@cutn.ac.in}
\affiliation{ 
Department of Physics, \\
School of Basic and Applied Sciences,\\
Central University of Tamil Nadu, Thiruvarur, \\
Tamil Nadu, India-610005
}%

\date{\today}

\begin{abstract}
In this paper, a novel method for the measurement of equilibrium contact angle of highly volatile binary liquids is proposed. The proposed method, which combines finite element method and energy equilibration, is able to calculate the solid-liquid contact area. The calculated solid-liquid contact area can then be used to estimate the equilibrium contact angle. Using the proposed approach, the contact angles of binary liquid droplet on a microgrooved and smooth polycarbonate substrate were calculated.  The proposed method can be an efficient tool for finding the contact angle of all liquids (both volatile and non-volatile).
\end{abstract}

\maketitle

%

\section{\label{sec:level1}Introduction  \lowercase{} }
Wettability becomes crucial for many industrial and scientific applications such as painting/coating \cite{prabhu, zhao, wang, sakai}, surface chemistry \cite{eral}, oil recovery \cite{prabhu, zhao, wang, sakai} and so on.  Wettability describes the balance of three inter-facial interactions namely, solid-liquid ($\gamma_{sl}$), liquid-vapor ($\gamma_{lv}$)and solid-vapor ($\gamma_{sv}$) inter-facial interactions \cite{good, gregory, hawking}. The balance between such three interactions are expressed by Young's equation \cite{young, youngdupre, malcolm}, 
\begin{equation}
\gamma_{lv} cos \theta = \gamma_{sv} - \gamma_{sl}
\label{1}
\end{equation}
where $\theta$ is the contact angle. 
Although the surface tension of the liquid can be measured experimentally with satisfactory accuracy, the solid-liquid inter-facial tension cannot be measured directly, and therefore the wettability is usually described by the contact angle, which is the angle formed between the solid-liquid interface and the liquid-vapor interface.  The contact angle is measured experimentally using a standard approach called sessile drop method, in which the camera focuses the liquid droplet placed over a solid substrate and the geometry of the droplet is used to obtain the contact angle \cite{bigelow}.\\

Apart from sessile droplet method, there are also other methods of measuring contact angle directly which includes "tilting plate" method \cite{macdougall, extrand}, captive bubble method, etc., \cite{mittal, taggart, wahlgren}. Although the measurement of contact angle from these methods is relatively straightforward, there are several issues that require attention while using volatile liquids: 1. The inherent inaccuracy of the direct measurement techniques and 2. Simultaneous variation in the contact area and contact angle of the liquid over the solid surfaces \cite{fowkes}.  There have been numerous studies reporting contact angles for a variety of liquids, binary systems, etc., In most such studies, the liquids used are less volatile and the binary system used contains water as one of its moiety. \\

In the present paper, an attempt is made to address the problem of measuring contact angle of highly volatile liquids.  Here, equilibrium droplet shape of highly volatile binary liquids on the horizontal smooth and constrained surfaces are simulated to obtain the contact angle of binary liquids on solid surfaces. This 3D-drop shape model is used to numerically analyze the shape, contact area and contact angle of the liquid droplets over the solid surface. The effect of variation of surface tension and surface roughness on the drop shape and apparent contact angle is examined. The liquids used in the present study includes water, methanol, ethanol and different concentrations of the ethanol-methanol binary system. The simulated results are validated with the experimentally obtained data. The present study shows that the proposed method can be an efficient tool for finding the contact angle of all liquids (both volatile and non-volatile).\\

The contact angle measurements are also useful in determining the size of the pores formed as a result of self-assembly of condensed liquid droplets. Since, pore size is a measure of diameter of a triple phase contact line. In the present study, an attempt is also made to compare the experimental pore size of the self-assembled droplet patterns with the simulated droplet size.  \\

\section{\label{sec:level1}Simulation Details}
The equilibrium shape of a liquid drop is achieved through energy minimization. In the present study, the finite element method and gradient descent method is used to evolve the surface toward minimal energy \cite{brakke1, brakke2}. The liquid droplet equilibration is achieved by minimizing various energies namely, surface tension, gravitational energy, etc., involved in the defined system.  The initial geometrical parameters, energies and constraints involved in the system are given as inputs and the program minimizes the total energy of the system by modifying the surface geometry according to the defined parameters and constraints. The droplet size obtained from simulated stable equilibrium shape of various liquid droplets over specific smooth and constrained surfaces are compared with those from experiments.  \\

\begin{figure}[h]
\centerline{\includegraphics[width=3in]{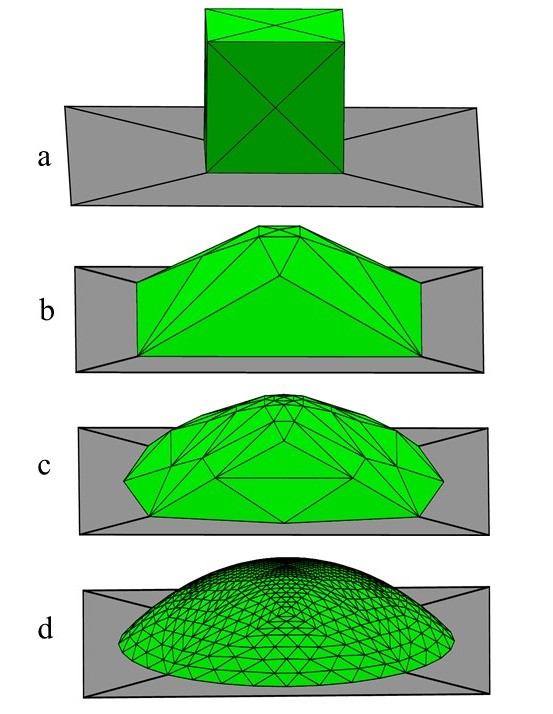}}
\caption{System evolving from initial geometry.}%
\label{initial}
\end{figure}

In the present work, the \textit{Surface Evolver} is used to simulate the drop of different liquids onto a smooth and constrained surface. The gravitational effect is negligible in the present case and hence do not influence the results significantly. Therefore, gravitational energy was not taken into account in our model.  The surface tension of the droplet and the inter-facial energy are specified in the data file.  The initial geometry and the shape after successive evolutions are shown in Figure~\ref{initial}. The free energy of the system is expressed as \cite{chen},
\begin{equation}
G/\gamma_{lv} = A_{lv} - \int\int \cos \theta dA
\label{44}
\end{equation}
and the contact angle ($\theta$) is defined by Young's equation \cite{young, youngdupre, malcolm}. Here in equation~\ref{44}, $\gamma_{lv}$ refers to surface tension of the liquid and $A_{lv}$ refers to the liquid-vapor contact area. The drop volume specified in this work is 9 $\mu l$.  The bottom face of the droplet is constrained to move and this boundary condition is considered to be responsible for obtaining the shape of the droplets. For the sake of convenience, the constraints are specified to the edges which defines the three phase contact line.  The successive refinements and steps concerning energy minimization computes the equilibrium shape of the system. Figure~\ref{initial} gives the illustration of the steps involved in the \textit{Surface Evolver} simulation.  \\

\section{\label{sec:level1}Results and Discussion}
\subsection{\label{sec:level2}Smooth surfaces}
\begin{figure}[!h]
\centerline{\includegraphics[width=3.5in]{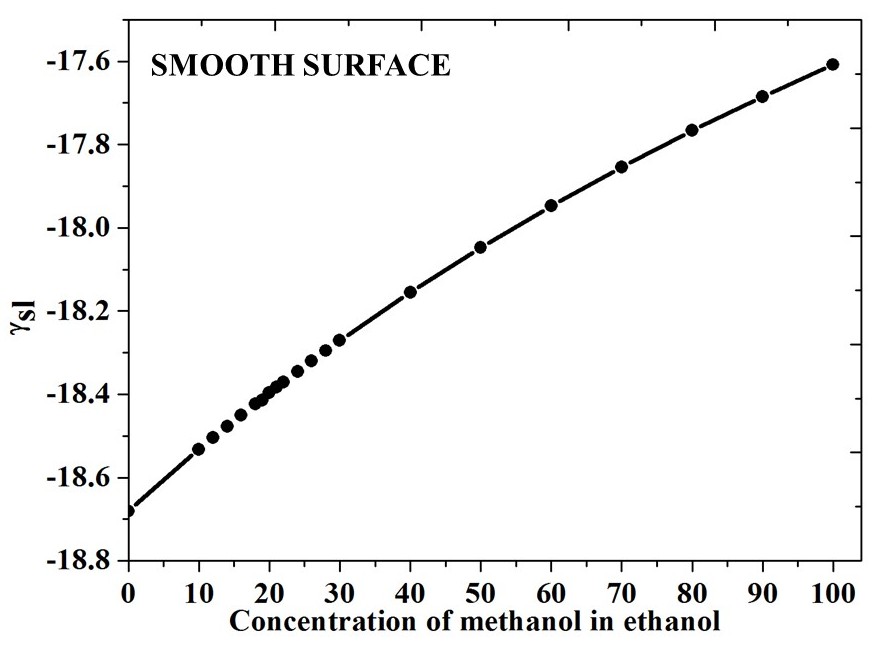}}
\caption{Variation of solid-liquid inter-facial tension of smooth surface for various concentration of methanol in ethanol.}%
\label{interfacial}
\end{figure}

Measuring the contact angle of the highly volatile liquid is experimentally difficult \cite{eral, mittal}.  Hence, in the present study, \textit{Surface Evolver} is used to model the wetting behavior of the ethanol-methanol binary liquid drop on the smooth surface. For this solid-liquid inter-facial tension calculated from experiments is used an input parameter instead of conventionally used contact angle.  The variation of solid-liquid inter-facial tension for various concentration of methanol is shown in Figure~\ref{interfacial}. \\
\begin{figure}[!h]
\centerline{\includegraphics[width=3.5in]{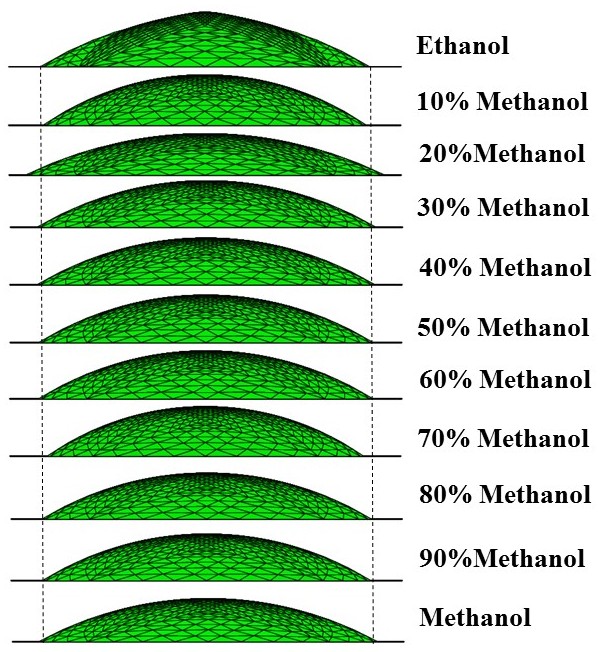}}
\caption{Variation of droplet diameter on smooth surfaces simulated using \textit{Surface Evolver}.}%
\label{11}
\end{figure}
The shape of the surfaces after the minimization of surface free energy are shown in Figure~\ref{11} and ~\ref{44}.  It is observed from the figure that the variation of concentration of binary liquid affects the width and height of the liquid drops on the smooth surface. Hence the shapes calculated from \textit{Surface Evolver} helps in understanding the variation of pores shapes with varying concentration of methanol \cite{nila}.  \\

\begin{figure}[h]
\centerline{\includegraphics[width=3.5in]{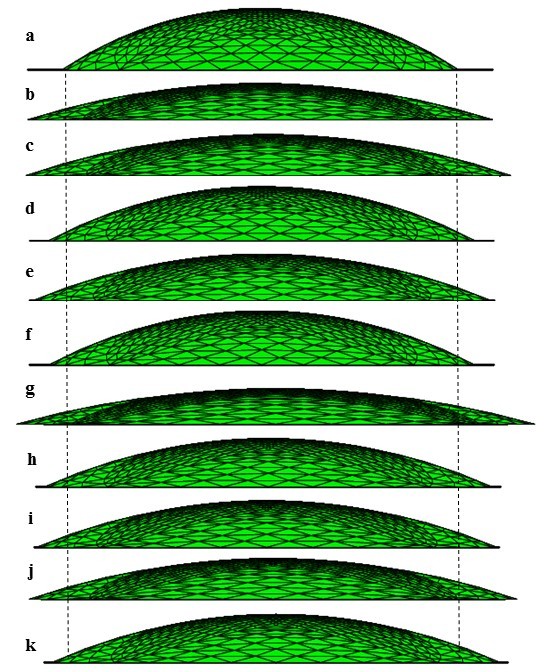}}
\caption{Variation of droplet diameter on smooth surfaces (From the top a)12\% b) 14\% c) 16\% d)18\% e) 19\% f) 20\% g) 21\% h)22\% i) 24\% j) 26\% k) 28\% of methanol in ethanol-methanol binary system).}%
\label{44}
\end{figure}
\begin{figure}[!h]
\centering{\includegraphics[width=4in]{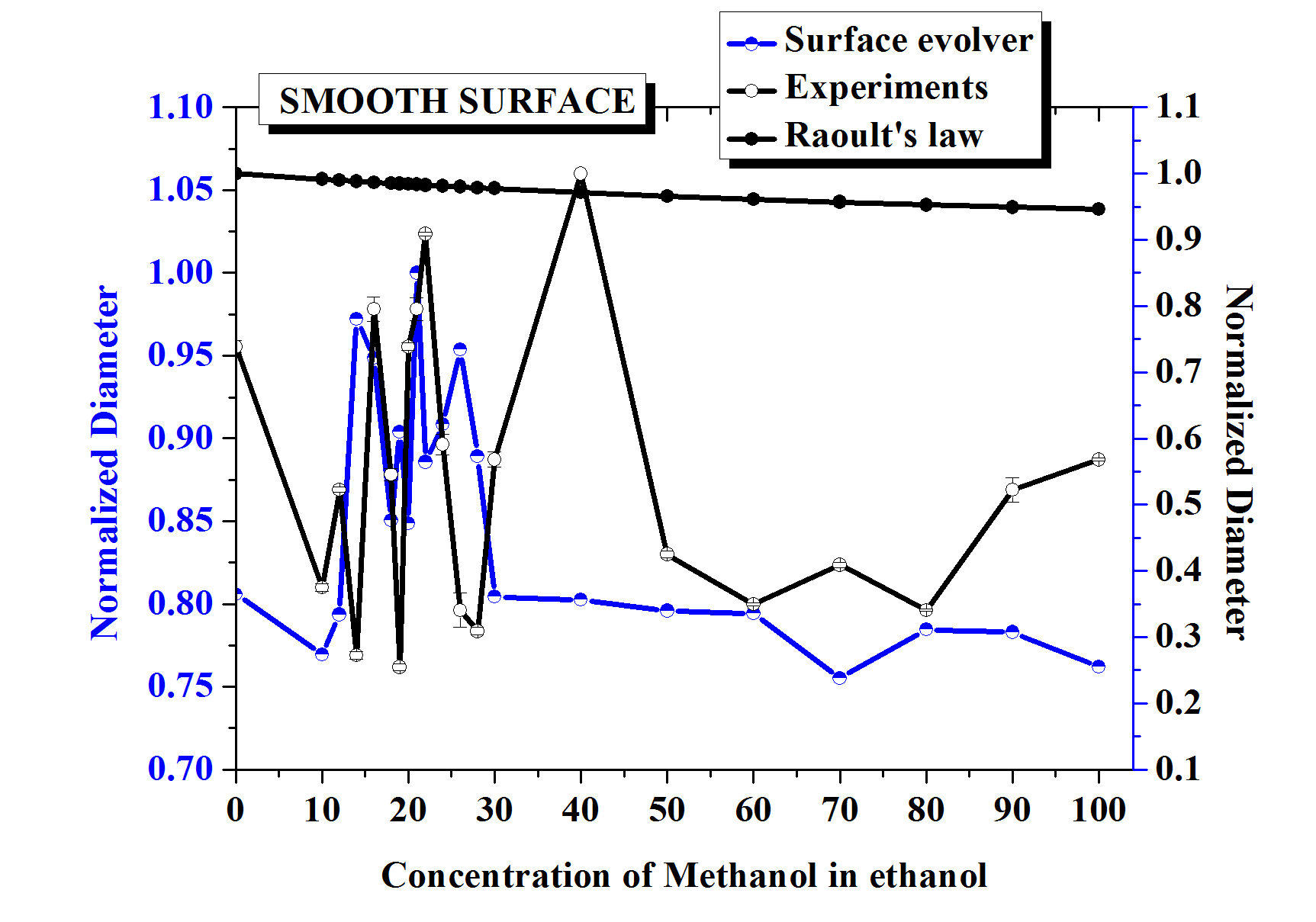}}
\caption{Comparison of normalized diameter on smooth surface calculated from simulation and experiments.}%
\label{normdia}
\end{figure}
\begin{figure}[!h]
\centering{\includegraphics[width=4in]{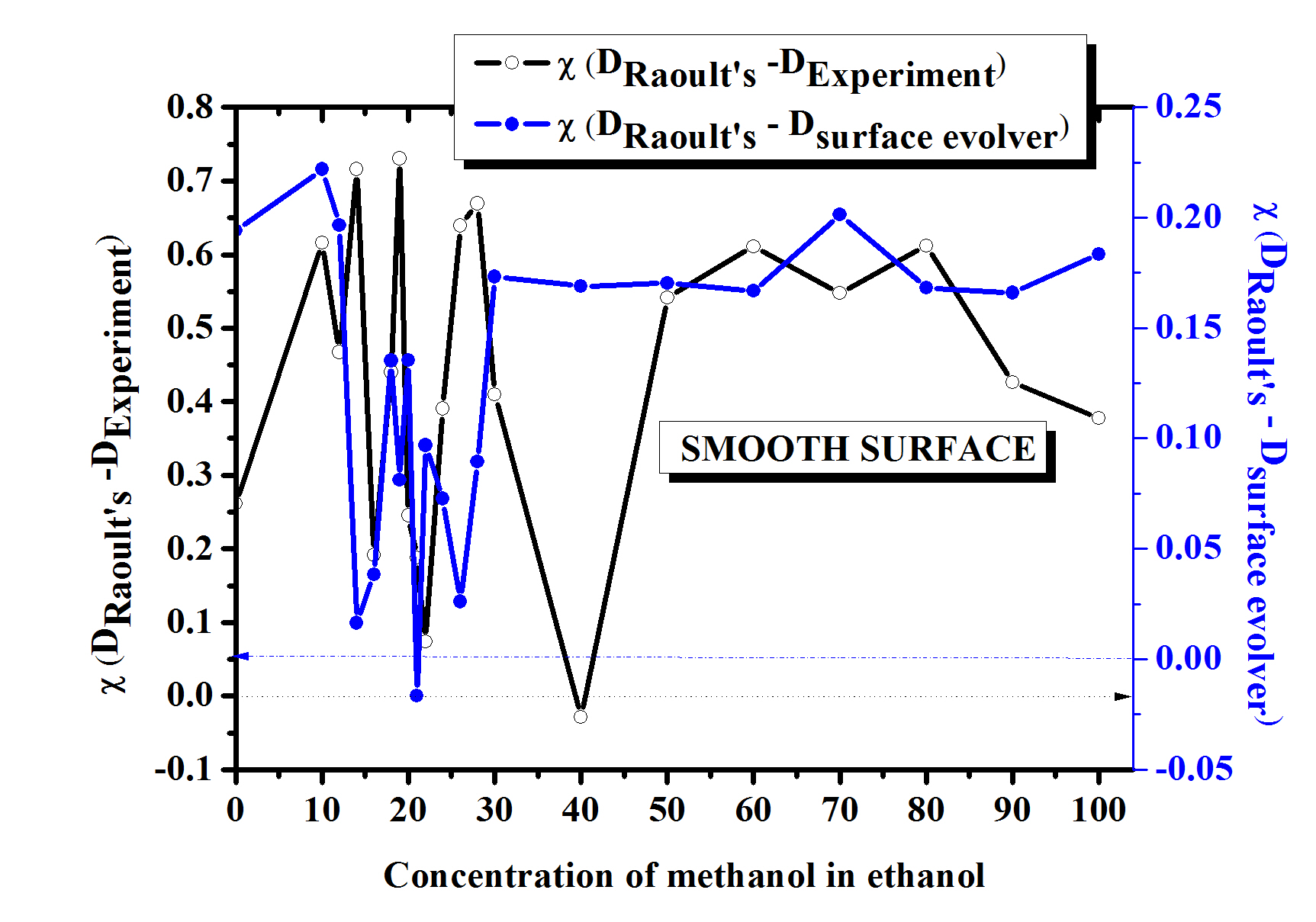}}
\caption{Deviation in experimental pore diameter and simulated droplet diameter from the ideal case (For smooth surface).}%
\label{chicomp}
\end{figure}
The solid-liquid inter-facial tension for all concentration of ethanol-methanol binary system is calculated following ideal Raoult's law \cite{mcquarrie, ebsmith} and using the obtained values, the drop shapes are simulated.  The results of \textit{Surface Evolver} simulation are compared with the experimentally observed profile of pore shapes. The simulation results show a similar trend as observed in experimental results \cite{nila}.  The complex variation at low concentration of methanol as seen in experiments is also observed in simulation results.  To further understand the correlation between the concentration of methanol and the pore size, diameter of the droplets are calculated.  From the obtained diameter of drops, normalized diameter is calculated. The normalized diameter versus concentration of methanol is also plotted (shown in Figure~\ref{normdia}). Comparison is made between the normalized diameter calculated from simulation and the normalized pore size obtained from experiments and is shown in Figure~\ref{normdia}. 

Further, the deviation in diameter obtained from experiment and simulation from the ideal case (diameter calculated from simulations using inter-facial tension calculated from Raoult's law) is calculated and is shown in Figure~\ref{chicomp}.  It is observed that except at the concentration of $21\%$ and $40\%$ of methanol, all other concentrations showed a positive value.  This concentration $21\%$ of methanol is the concentration where the complexity in intermolecular interactions is observed \cite{nilaspec} and $40\%$ of methanol is the concentration, where the presence of strong hydrogen bonds were reported \cite{alvarez, nilaspec}. \\

\begin{figure}[h]
\centering{\includegraphics[width=3.5in]{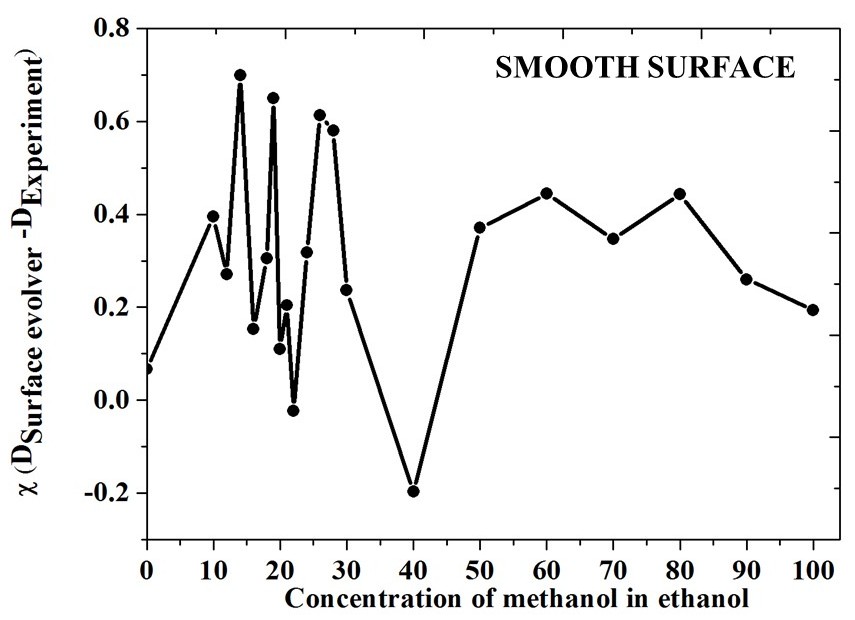}}
\caption{Difference between diameter (on smooth surface) obtained from simulation and experiment.}%
\label{chi}
\end{figure}

\begin{figure}[h]
\centering{\includegraphics[width=4in]{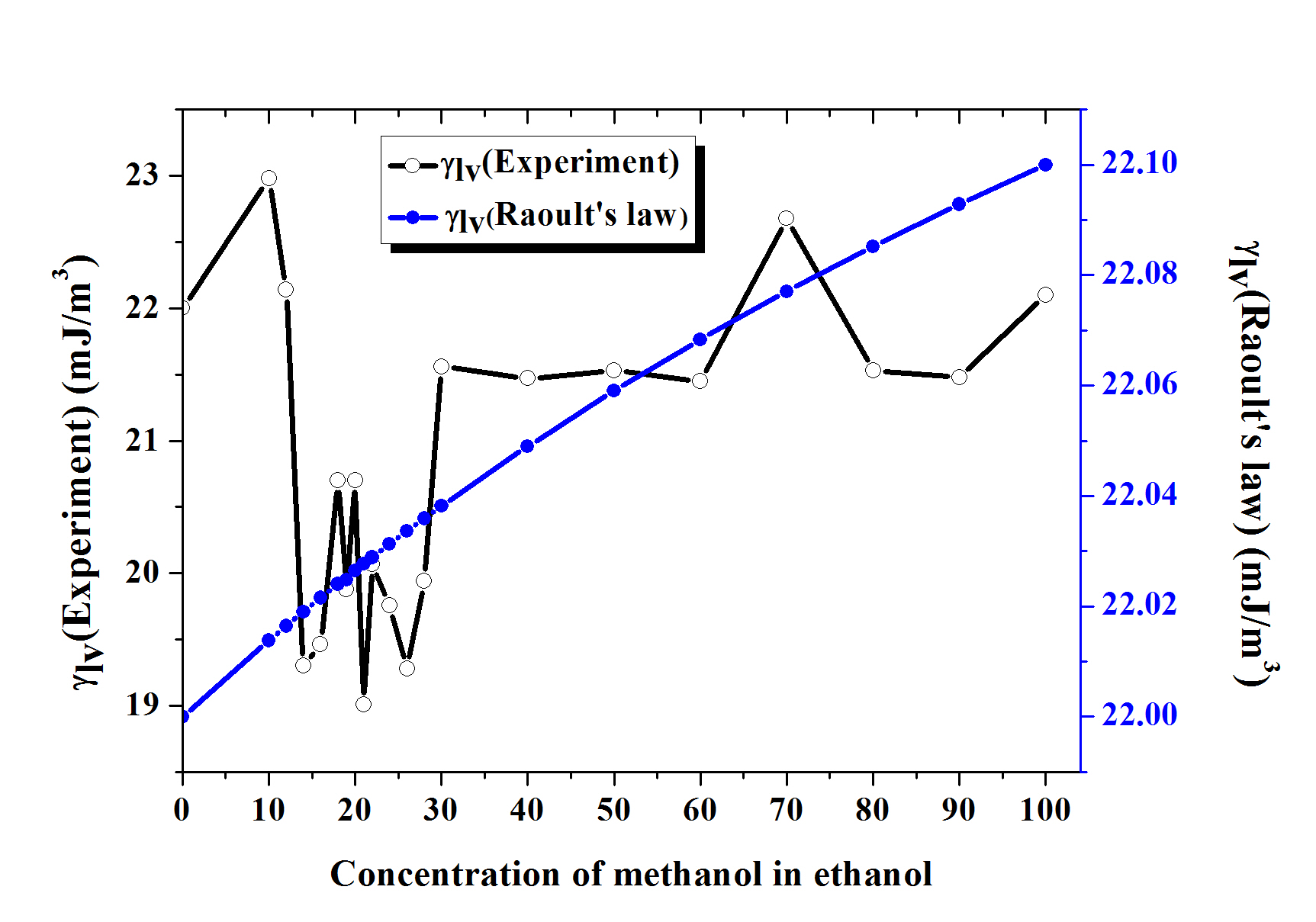}}
\caption{Surface tension of liquids obtained from experiments and Raoult's law.}%
\label{gamma}
\end{figure}

The difference between the simulation and experimental diameter is also calculated and is shown in Figure~\ref{chi}.  Positive deviation is observed for all concentration of methanol  except at $21\%$ of methanol. This shows that the simulated results are in good agreement with the experimental data i.e., the relative error between experimental and simulated values are small. The negative value at $21\%$ occurs in the region where the complex behavior in inter-molecular interactions is observed. For the sake of comparison, the liquid-vapor inter-facial tension obtained from experiments and calculated from Raoult's law is plotted and shown in Figure~\ref{gamma}.  It is clear from the figure that the ethanol-methanol binary system is far from ideal. \\
   
\newpage{}
\begin{figure}[!h]
\centering{\includegraphics[width=3.5in]{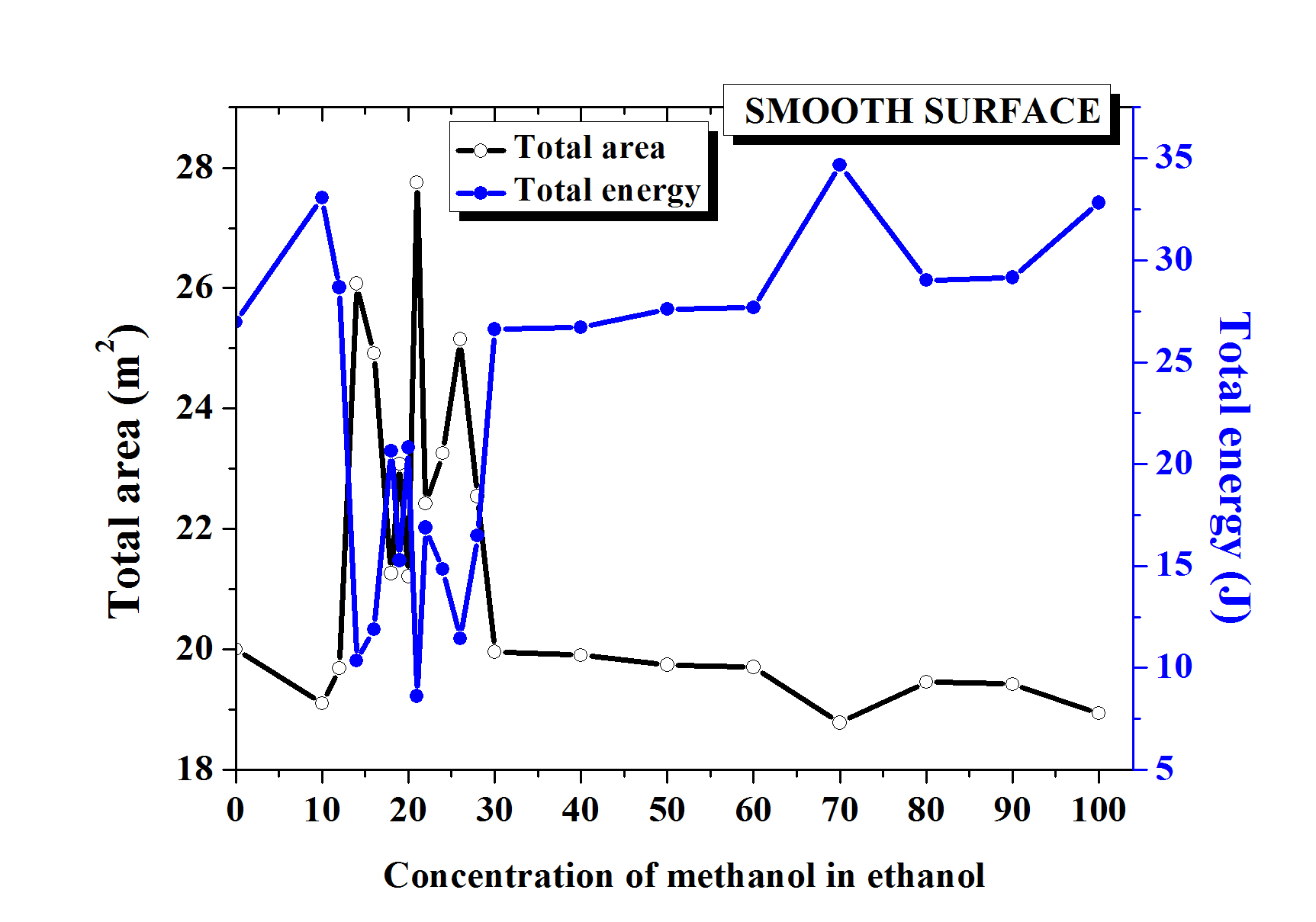}}
\caption{Variation of total energy and total area of the system (liquid droplet on smooth surface) for various concentration of methanol in ethanol.}%
\label{energy}
\end{figure}
The total energy and total area of the system for various concentration of methanol are calculated from the simulations and are shown in Figure~\ref{energy}.  The total energy and total area show a similar trend as seen from experimental result at the lower concentration of methanol in ethanol-methanol binary system.  \\

Hence in conclusion, the \textit{Surface Evolver} results is found to corroborate the experimental findings. \textit{Surface Evolver} is proved to be a tool for calculating the contact angle of highly volatile liquids for which experimental measurement of contact angle is difficult.\\

\subsection{\label{sec:level2}Constrained surfaces}
\begin{figure}[h]
\centerline{\includegraphics[width=4in]{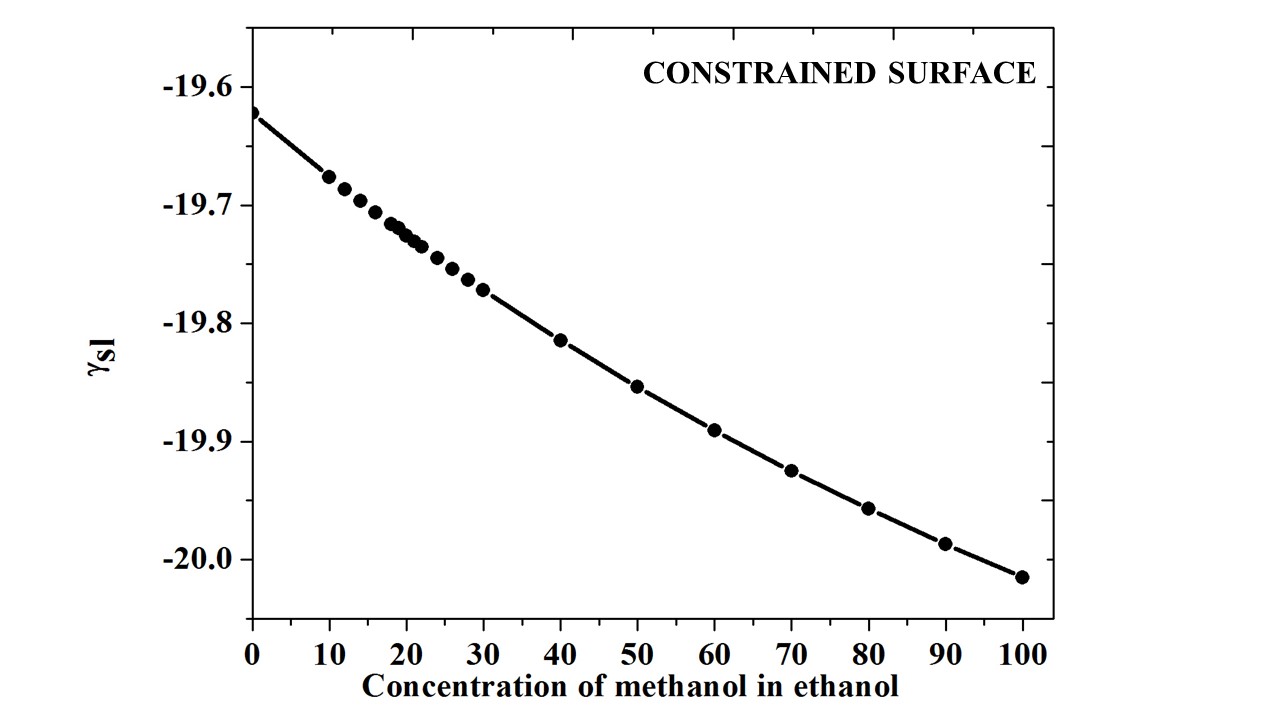}}
\caption{Variation of solid-liquid inter-facial tension of constrained surfaces with various concentration of methanol in ethanol.}%
\label{interfacialconst}
\end{figure}

The solid-liquid inter-facial tension for constrained surfaces are calculated from experiments  and used as an input parameter instead of conventionally used contact angle.  The variation of solid-liquid inter-facial tension for various concentration of methanol is shown in Figure~\ref{interfacialconst}. \\

\begin{figure}[h]
\centerline{\includegraphics[width=3.5in]{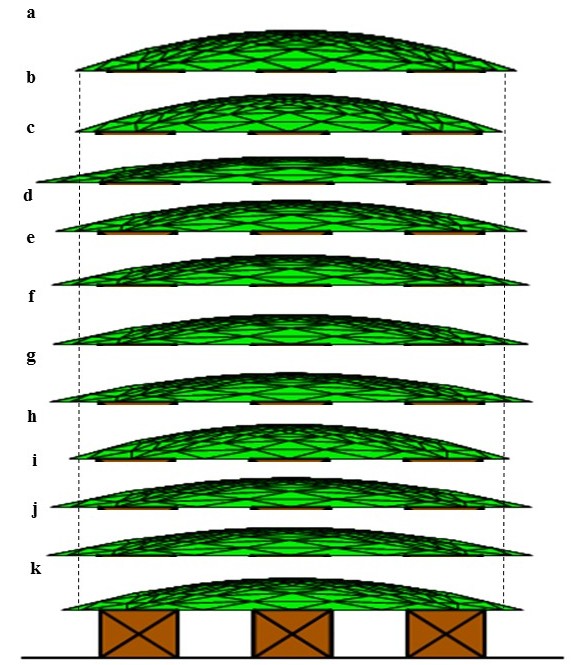}}
\caption{Variation of droplet diameter on constrained surfaces (From the top a) 0\% b) 10\% c) 20\% d) 30\% e) 40\% f) 50\% g) 60\% h) 70\% i) 80\% j) 90\% k) 100\% of methanol in ethanol-methanol binary system).}%
\label{dropconsvaria}
\end{figure}

\begin{figure}[h]
\centerline{\includegraphics[width=3.5in]{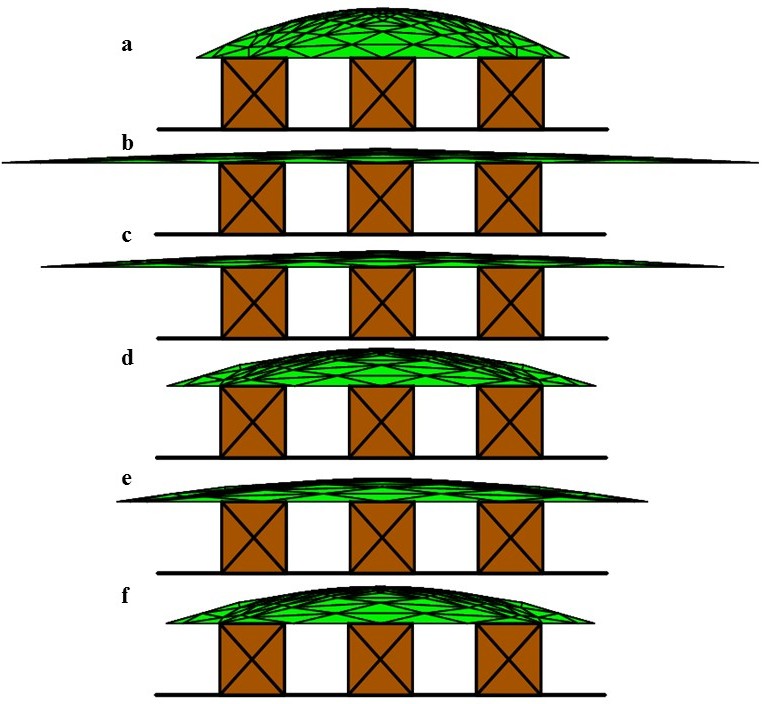}}
\caption{Variation of droplet diameter on constrained surfaces (From the top a) 12\% b) 14\% c) 16\% d) 18\% e) 19\% f) 20\% g) 21\%  of methanol in ethanol-methanol binary system).}%
\label{dropfinevaria}
\end{figure}

The shape of the droplet on constrained surfaces after the energy minimization are shown in Figure~\ref{dropconsvaria} and ~\ref{dropfinevaria}.  It is observed from the figure that the variation of concentration of binary liquid affects the width and height of the liquid drops on the constrained surface. Hence the shapes calculated from \textit{Surface Evolver} helps in understanding the variation of pores shapes with varying concentration of methanol. It is also observed that the constraints on the surface has significant influence in the wetting behavior of the surface.  \\

\begin{figure}[h]
\centerline{\includegraphics[width=3.5in]{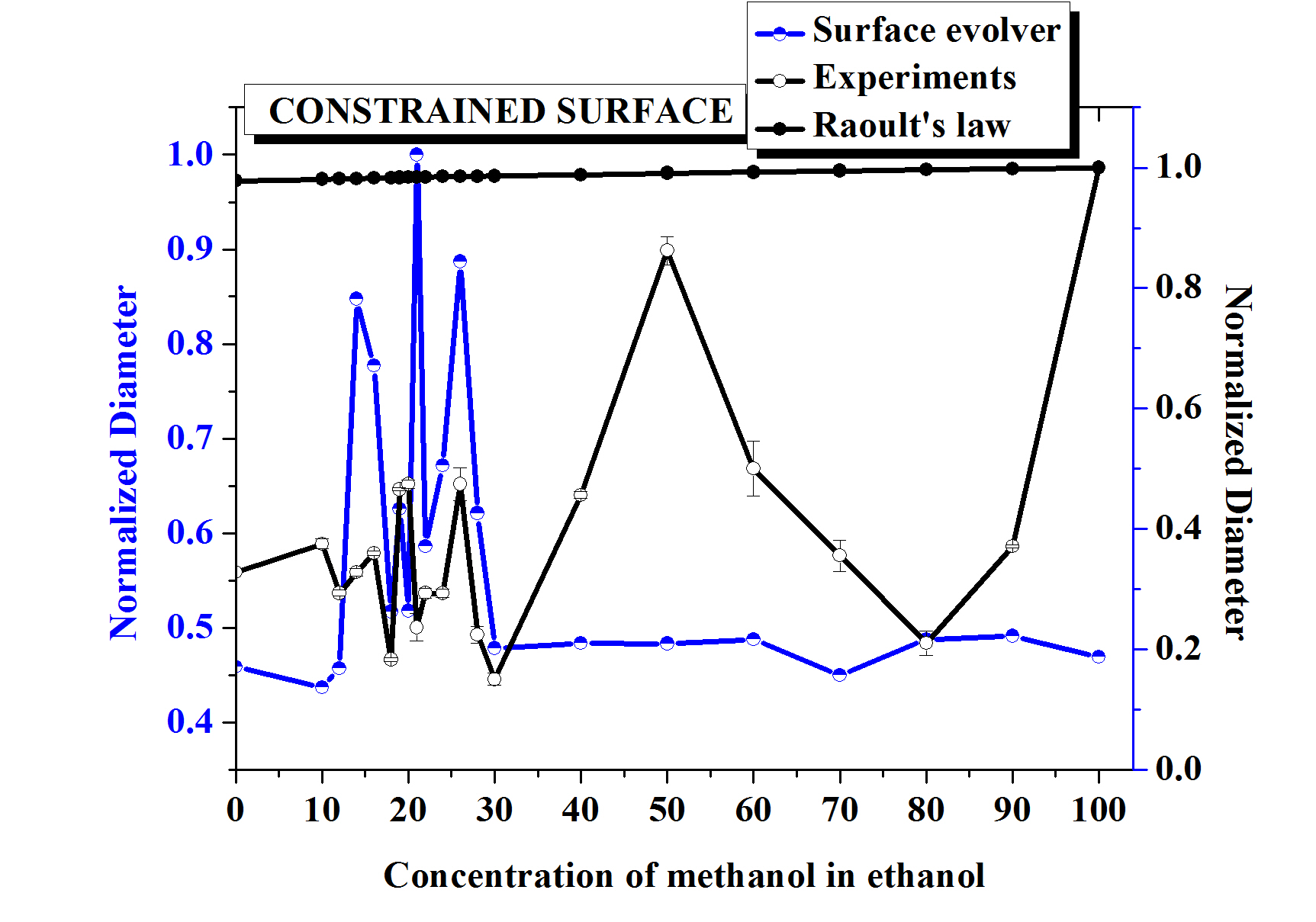}}
\caption{Normalized diameter of droplet on constrained surfaces obtained from simulation and experiment.}%
\label{normdiaconst}
\end{figure}

Similar to smooth surfaces, the solid-liquid inter-facial tension for all concentration of ethanol-methanol binary system is calculated following ideal Raoult's law and using the obtained values the drop shapes are simulated.  The results of \textit{Surface Evolver} simulation are compared with the experimentally observed profile of pore shapes on constrained surfaces. The simulation results show a similar trend as observed in experimental results.  The complex variation at low concentration of methanol is also observed in the simulation results.  To further understand the correlation between the concentration of methanol and the pore size, diameter of the droplets are calculated.  From the obtained diameter of drops, normalized diameter is calculated. The normalized diameter versus concentration of methanol is also plotted (shown in Figure~\ref{normdiaconst}). Comparison is made between the normalized diameter calculated from simulation and the normalized pore size obtained from experiments and is shown in Figure~\ref{normdiaconst}. 
\begin{figure}[!h]
\centerline{\includegraphics[width=3.5in]{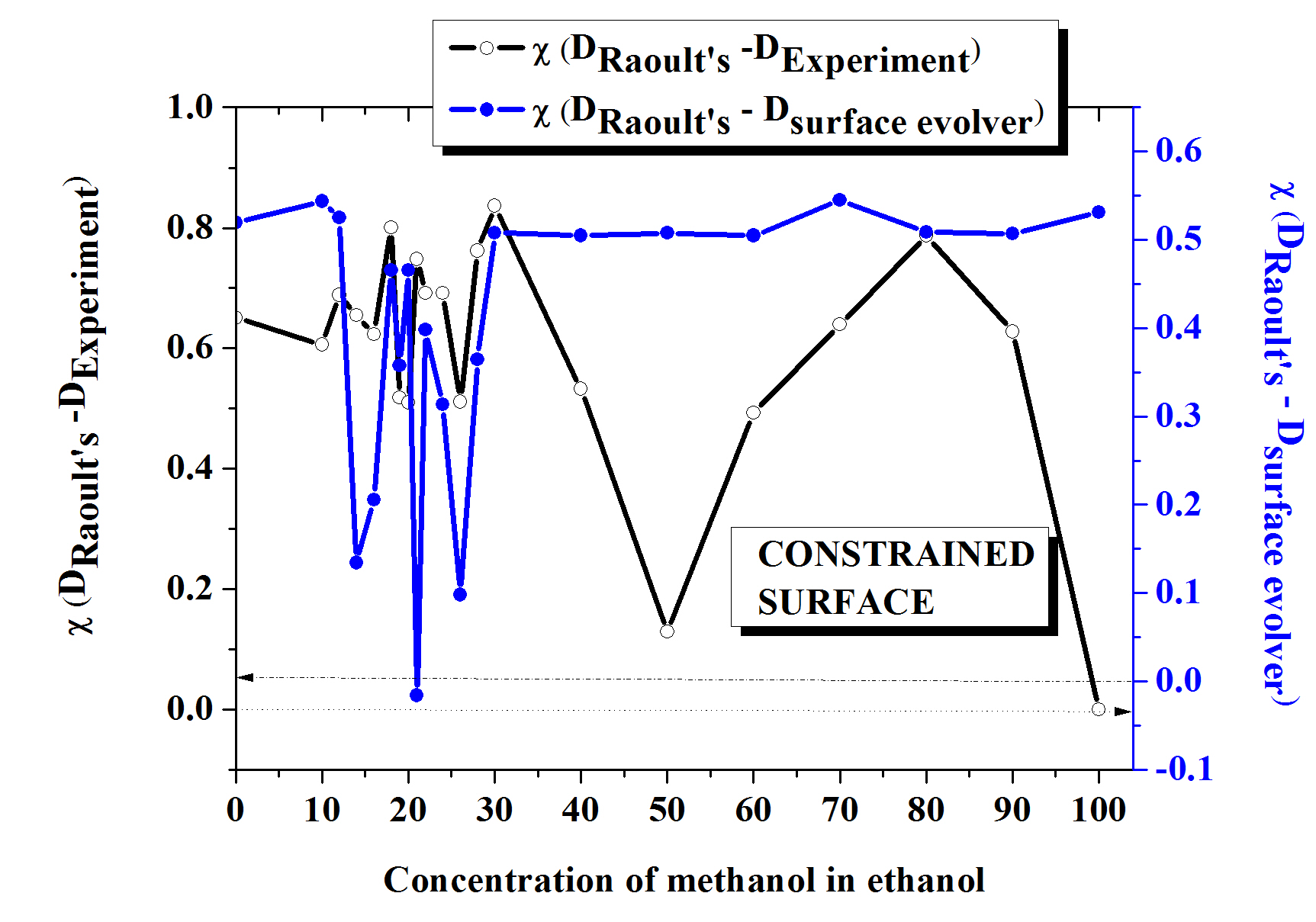}}
\caption{Deviation in experimental pore diameter and simulated droplet diameter from ideal case (constrained surfaces).}%
\label{chicompconst}
\end{figure}
Further, the deviation in diameter obtained from experiment and simulation from the ideal case (diameter calculated from simulations using inter-facial tension calculated from Raoult's law) is calculated and is shown in Figure~\ref{chicompconst}. Positive deviation is observed for all concentration of methanol  except at $21\%$ of methanol and $100\%$, where the complexity in properties is observed.  At certain concentrations, the datas obtained from experiments and simulations doesnot correlate.  This is due to the presence of chloroform in the experiments which is not taken into account in simulations.  \\

\begin{figure}[h]
\centerline{\includegraphics[width=3.5in]{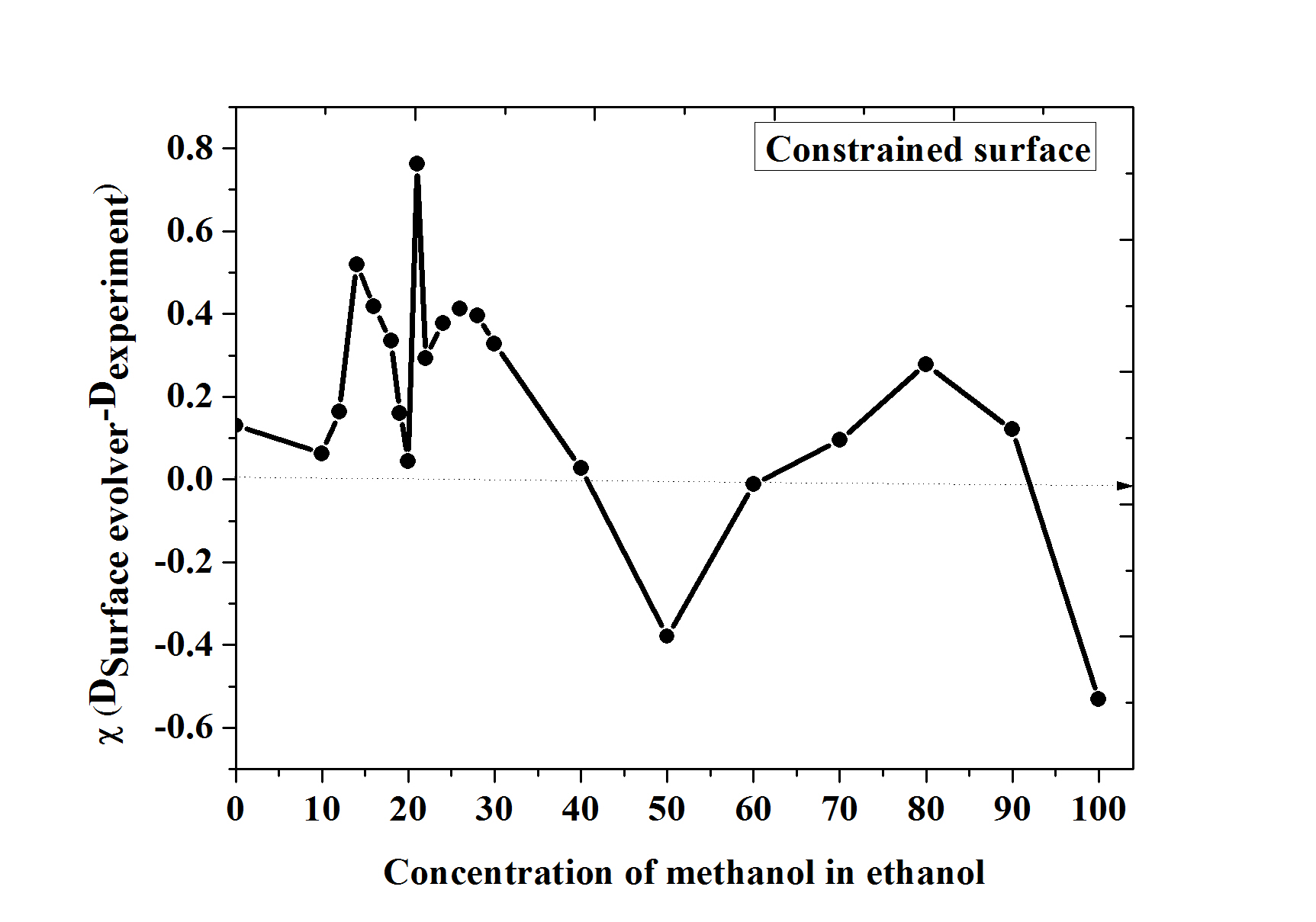}}
\caption{Difference between the simulated and experimental diameter for constrained surfaces.}%
\label{chiconst}
\end{figure}

The difference between the simulation and experimental diameter for constrained surfaces is also calculated and is shown in Figure~\ref{chiconst}.  It is observed that the deviations is observed to be positive for low concentration of methanol. The higher concentration shows a negative deviation, which may be due to the possibility of complex interactions as observed in \cite{nila, nilaspec}. For the sake of comparison, the liquid-vapor inter-facial tension obtained from experiments and calculated from Raoult's law is plotted and shown in Figure~\ref{gamma}.  It is clear from the figure that the ethanol-methanol binary system is far from ideal. \\
\begin{figure}[!h]
\centerline{\includegraphics[width=3.5in]{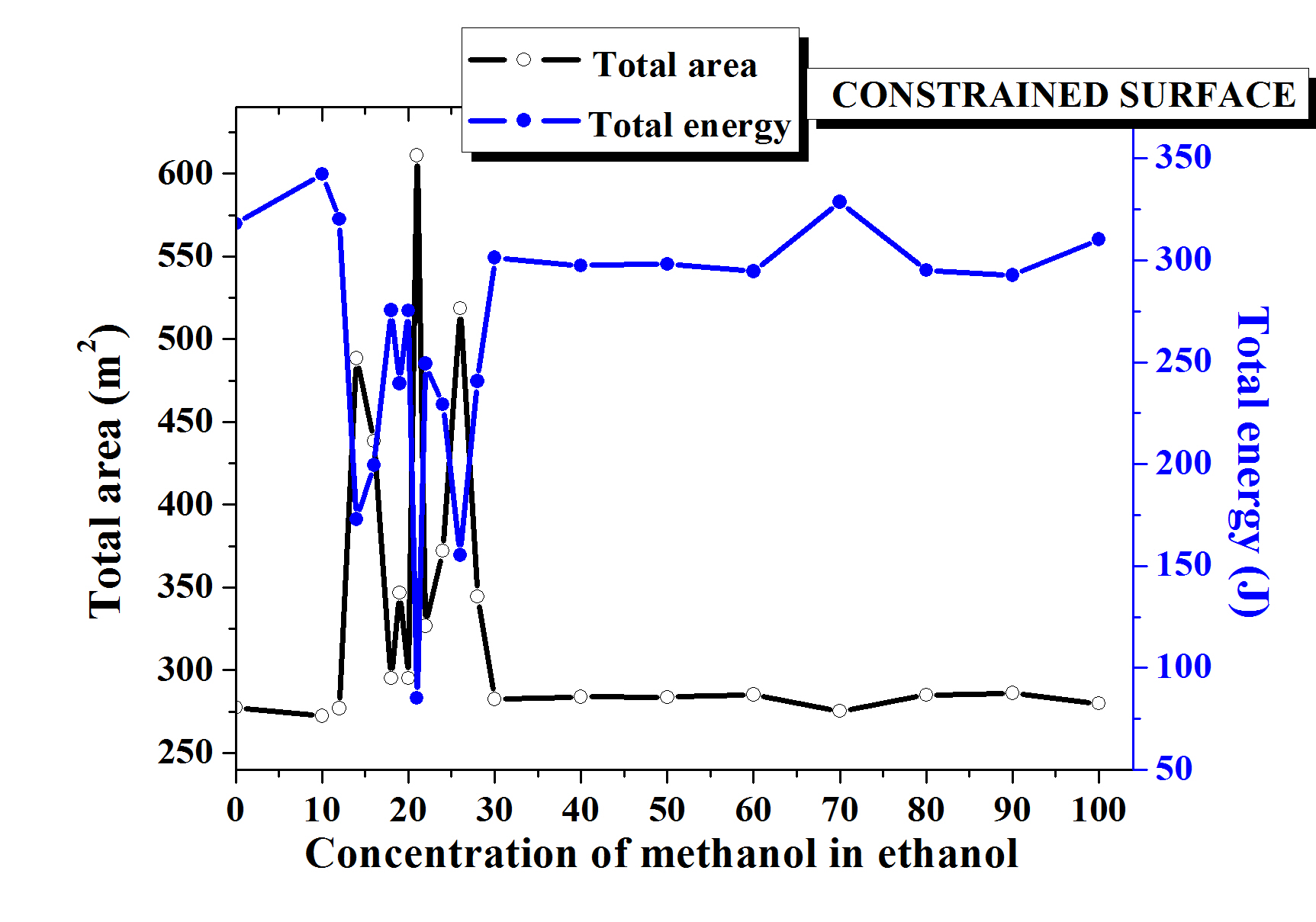}}
\caption{Variation of total area and total energy of the system (liquid droplet on constrained surfaces) .}%
\label{energyconst}
\end{figure}
The total energy and total area of the constrained system for various concentration of methanol are calculated from the simulations and are shown in Figure~\ref{energyconst}.  The total energy and total area show a similar trend as seen from experimental result at the lower concentration of methanol in ethanol-methanol binary system.  \\

Similar to the simulations obtained for smooth surfaces,  the \textit{Surface Evolver} results of constrained surfaces is found to corroborate the experimental findings. \textit{Surface Evolver} is proved to be a tool for calculating the contact angle of highly volatile liquids for which experimental measurement of contact angle is difficult.\\ 

\section{\label{sec:level1}Conclusion}
 The contact angle of highly volatile liquid droplets on horizontal smooth and constrained surfaces are calculated. This is performed by simulating the equilibrium liquid droplet shape on the specified surfaces and numerically analyzing the contact area and contact angle of the liquid droplets on the surfaces.  The effect of variation of surface tension and surface roughness on the drop shape and apparent contact angle are examined through the simulating the droplets of water, methanol, ethanol and various concentration of ethanol-methanol binary system.  The simulated results are validated with the experimentally obtained data. The present study shows that the \textit{Surface Evolver} can be an efficient tool for finding the contact angle of all liquids (both volatile and non-volatile).\\


\begin{thebibliography}{100}
\bibitem{prabhu}
K.N. Prabhu, P. Fernandes, G. Kumar, Mater. Des. 2, 297 (2009)
\bibitem{zhao}
X. Zhao, M.J. Blunta, J.J. Yao, Pet. Sci. Technol. Eng. 71, 169 (2010)
\bibitem{wang}
Y.Q.Wang, H.F. Yang, Q.G. Hang, L. Fang, S.R. Ge, Adv.Mater. Res. 154-155, 1019 (2010)
\bibitem{sakai}
M. Sakai, T. Yanagisawa, A. Nakajima, Y. Kameshima, K. Okada, Langmuir 25, 13 (2009)
\bibitem{eral}
H. B. Eral, D. J. C. M. 't Mannetje, and J. M. Oh., Colloid Polym. Sci., 291, 2:247-260, (2013).
\bibitem{good}
Robert J. Good, Journal of Adhesion Science and Technology, 6:12, 1269-1302, (1992).
\bibitem{gregory}
Gregory J. Merchant, Physics of Fluids A: Fluid Dynamics, 4, 477 (1992).

\bibitem{hawking}
L. M. Hawking,  Physics of Fluids, 7, 2950 (1995).

\bibitem{young}
T. Young, Philos. Trans. R. Soc. Lond. 95, 65 (1805)

\bibitem{youngdupre}
A. M. Dupre, Theorie Mechanique de la Chaleur, Gauthier-Villars, Paris, 369, (1869).

\bibitem{malcolm}
M. E. Schrader, Langmuir, 11, 3585-3589, (1995).

\bibitem{bigelow}
W.C. Bigelow, D.L. Pickett, W.A.J. Zisman, Coll. Sci. 1, 513 (1946)

\bibitem{macdougall}
G. MacDougall, C. Ockrent, Proc. R. Soc. 180A, 151 (1942)

\bibitem{extrand}
C.W. Extrand, Y. Kumagai, J. Colloid Interface Sci. 184, 191 (1996)

\bibitem{mittal}
K. L. Mittal and R. Jaiswal, Particle Adhesion and Removal, Scrievener publishers, (2014).

\bibitem{taggart}
A.F. Taggart, T.C. Taylor, C.R. Ince, Trans. Am. Inst. Min. Metall. Pet. Eng. 87, 285 (1930).


\bibitem{wahlgren}
W. Zhang, M. Wahlgren, B. Sivik, Desalination, 72, 263 (1989)

\bibitem{fowkes}
Fowkes, Contact Angle, Wettability, and Adhesion Advances in Chemistry; American Chemical Society: Washington, DC, (1964)

\bibitem{brakke1}
K. A. Brakke, The Surface Evolver, Exp. Math. 1, 141, (1992)

\bibitem{brakke2}
K. A. Brakke, 2008, Surface Evolver manual: version 2.30,  http://www.susqu.edu/brakke/aux/downloads/manual230.pdf

\bibitem{chen}
Y Chen, B He, J Lee, N A Patankar, Journal of Colloid and Interface Science, 281, 458-464, (2005)

\bibitem{nila}
K. Nilavarasi and V. Madhurima, Eur. Phys. J. E,  41: 82, (2018).

\bibitem{mcquarrie}
D. A. McQuarrie, J. D. Simon, University Science Books, (1997).

\bibitem{ebsmith}
E. B. Smith, Clarendon Press, Oxford, (1993).

\bibitem{nilaspec}
K. Nilavarasi, Thejus R Kartha and V. Madhurima, Spectrochimica Acta Part A: Molecular and Biomolecular Spectroscopy, 188, 301-310, 2018. 

\bibitem{alvarez}
V. H. Alvarez, S. Mattedi, M. Iglesias, R. Gonalez-Olmos and J. M. Resa, Physics and Chemistry of Liquids, 49:1, 52-71, (2011).

\bibdata{all}
\end{thebibliography}

\end{document}